\documentclass[aps,twocolumn,showpacs,floatfix]{revtex4}
\usepackage{dcolumn}
\usepackage{graphicx}
\usepackage{amsfonts}
\usepackage{amsmath,bm,amssymb}

\input{tcilatex}
\begin{document}

\title{}

\textbf{Han and Savrasov's Reply:} In Singh's comment \cite{Singh} two
statements are made regarding our Letter \cite{Letter}: (a) Local Density
Approximation (LDA) based calculations of doped Fe$_{1+x}$Te do not show the
Fermi surface nesting at ($\pi $,0) when $\sim $0.5 electrons is added per
Fe (upward energy shift by $\sim $0.4 eV) and show it at only when $\sim $1
electron is added per Fe (upward energy shift by $\sim $0.7 eV) using rigid
band approximation (RBA); (b) Coherent Potential Approximation (CPA)\
calculations of doped FeTe do not show nesting at ($\pi $,0) at those
dopings.

In regard to (a) we appreciate this discrepancy being pointed out to us and indeed
discovered \textquotedblleft a factor of two\textquotedblright\ error in
extracting the doping in our original publication \cite{Letter}.
However, as we discuss below, the numbers deduced from
LDA cannot be trusted due to correlation effects; therefore the detailed
quantitative analysis in Singh's comment cannot be taken seriously into
account for this system. We have already pointed out earlier \cite{Yin} that
LDA has an unprecedented error in determination of the z structural position
of anion in pnictides. This, in particular, may lead to a 20\% uncertainty in
determining the density of states near the Fermi level. A different and
possibly much larger source of error lies in the fact that there is a
non--trivial frequency and orbital dependent self--energy correction to LDA.
For example, previous LDA+Dynamical Mean Field Theory (DMFT) calculations 
\cite{Haule} found band mass enhancements $m/m_{LDA\text{ }}$to be between 2 
\cite{Anisimov} to 5 \cite{Haule}. Angle resolved photoemission (ARPES)\
experiments show $m/m_{LDA}\approx 2$ for pnictides \cite{ARPES} while the
situation is controversial for chalcogenides. Xia et al. \cite{Xia} reported 
$m/m_{LDA}$ in Fe$_{1+x}$Te similar to pnictides which, however, contradicts
with the specific heat coefficient \cite{SpecificHeat}
by a factor of 4. A recent publication \cite{ARPES-FeTe} finds much larger
and anisotropic mass enhancements $\approx 6-23$ in FeSe$_{0.42}$Te$_{0.58}$
which also agrees much better with the measured Sommerfeld coefficient of
Ref. \cite{SpecificHeat}. On top of that the order of the bands near the $%
\Gamma $ point seen by this ARPES experiment is in accord with LDA
calculations for FeSe but not for FeTe. The fact that chalcogenides are more
correlated than pnictides is also evident from comparative analysis of low
energy model Hamiltonians derived by a first principle electronic structure
calculation.\cite{Takashi}.

The effect of correlation in the vicinity of the Fermi energy leads to the
self--energy correction for the electron in the form $\Sigma _{\alpha
}(\omega )=\Sigma _{\alpha }(0)+\omega (1-z_{\alpha }^{-1})$ where $%
z_{\alpha }$ is the quasiparticle residue$.$ Therefore it is clear that the
precise value of the energy shift necessary to change the topology of the
Fermi surface depends on the mass enhancement and the value of 0.7 eV
extracted from LDA is incorrect. For example, a mass enhancement of 10 would
assume an upward energy shift by merely 0.07 eV to produce ($\pi $,0)
nesting. A different consideration applies to the error in a number of
electrons because isotropic mass enhancement alone will not affect the level
of doping needed to change the Fermi surface due to Lattinger's theorem. In
this regard, if the mass enhancement is fairly band independent as it was
seen in LDA+DMFT calculations for pnictides, the same value of $\sim $1
electron per Fe would be needed. However, the situation may change since
strong orbital dependence in the mass enhancement was recently reported
together with the indication that some orbital dependent shift $\Sigma
_{\alpha }(0)$ is needed to account for the correct order of the bands\cite%
{ARPES-FeTe}. Under this circumstance, the Lattinger theorem does no longer
hold and the value of doping needed to switch to the ($\pi $,0) nesting will
be smaller.

In regard to Singh's comment (b) that CPA does not see the nesting: it is
interesting but needs further study in its relevance to the physics of these
systems. We would like to point out two things: first, several ARPES studies
of doped Fe superconductors appeared in the literature; they may not see a
disorder oriented broadening of the Fermi surfaces as reported by this CPA
calculation. Second, a recent supercell study of doping dependent band
structures in pnictides \cite{Sashi} concluded that virtual crystal
approximation (VCA) is adequate to model the doping. We have performed our
own VCA calculations assuming a uniform positive background compensating for
extra electrons and verified that while there are important changes in band
dispersions in the vicinity of Fermi level, the topology of the Fermi
surface does follow the rigid band analysis of our original work and ($\pi $%
,0) nesting is still present.

To summarize, inclusion of correlation effects affects quantitatively the
agreement with experiment as far as the value of energy shift and the level
of doping is concerned, and our original statement that nesting at ($\pi $
,0) can be responsible for magnetic behavior of FeTe is hereby reinstated.

\end{document}